\documentclass[letterpaper,journal,twoside,web,onecolumn,12pt]{IEEEtran}
\usepackage{cite}
\usepackage{amsmath,amssymb,amsfonts}
\usepackage{algorithmic}
\usepackage{graphicx}
\usepackage{textcomp}
\usepackage{amsthm} 

\newtheorem{lem}{Lemma}

\newtheorem{thm}{Theorem}

\newtheorem{rem}{Remark}

\title{MIMO Super-Twisting Controller using a passivity-based design}

\author{Juan F. Garc\'{i}a-Mathey, Jaime A.~Moreno \IEEEmembership{Member, IEEE}   
\thanks{
Instituto de Ingenier\'{i}a, Universidad Nacional Aut\'{o}noma de M\'{e}xico (UNAM), Ciudad de Mexico, Mexico. JMorenoP@ii.unam.mx}
\thanks{This work was supported by PAPIIT-UNAM, project IN102121 and by CONACyT, CVU 1083638. }
}

\begin{document}
\maketitle
\thispagestyle{empty}

\begin{abstract}
A novel MIMO homogeneous Super-Twisting Algorithm is proposed in this paper for nonlinear systems with relative degree one, having a time and state-varying uncertain control matrix. The uncertainty is represented by a constant but unknown left matrix factor. Sufficient conditions for stability with full-matrix control gains are established, in contrast to the usual scalar gains. For this a smooth Lyapunov function, based on a passivity interpretation, is used. Moreover, continuous and homogeneous approximations of the classical discontinuous Super-Twisting algorithm are obtained, using a unified analysis method. 
\end{abstract}


\section{Introduction}
\label{sec:Introduction}
%

\IEEEPARstart{T}{he} super-twisting algorithm is a classic
result of Sliding-Mode Control\cite{FriLev02,ShtEdw14}, used for control \cite{Lev93} and differentiaton \cite{Lev98}. Its design has been performed by geometric methods \cite{Lev93}, using Lyapunov functions \cite{MorOso12,SeeHor18,SeeHor19} and also frequency domain methods \cite{PerFri19}. %
It is a popular control algorithm because of its unique features
and advantages for systems of relative degree $1$: 
it can compensate matched Lipschitz perturbations
and uncertainties; it forces the output and its derivative to zero
in finite-time, while only requiring knowledge of the output; and it
generates a continuous control signal \cite{FriLev02,ShtEdw14}.

Although the scalar algorithm is rather well-understood, the MIMO versions are in their infancy. For control purposes, 
if the plant's control matrix is known, a preliminary decoupling allows to perform the control design using independent scalar systems. This is done e.g. in \cite{MerMor21}, using the Implicit Lyapunov Function approach. However, in the case that the control matrix is uncertain or that the decoupling is not desired, 
truly multivariable analysis and design methods are imperative.

At present, there exist only few MIMO versions of the Super-Twisting algorithm. \cite{NagEdw14} is the first reported in the literature. Although it assumes that the control matrix is known, and a decoupling can be performed, the control terms are of the quasi-continuous (or unitary) type, so that they induce a coupling in all the control channels. \cite{NagEdw14} uses a non smooth Lyapunov function, extending the one proposed in \cite{MorOso08}, and includes also an extra linear term, that allows to compensate for uncertainties and perturbations growing with the state of the plant. \cite{LopMor15,LopMor19} provide a further generalization. Assuming  a known control matrix, 
two different schemes are proposed: one with decoupled control terms, 
i.e. a set of SISO algorithms, and another with quasi-continuous control terms, as in \cite{NagEdw14}. Varying gains and extra terms, more general than linear ones, are also allowed, that permit an acceleration of the convergence and the compensation of state growing uncertainties and/or perturbations. 
A characterization and comparison of the kind of convergence provided by the 
discontinuous and the quasi-continuous control terms is presented. A non-smooth Lyapunov function is used for analysis and design. 

\cite{KamCha15} combines the MIMO Super-Twisting proposed in \cite{NagEdw14} with Integral Sliding-Mode Control, to reduce the chattering effect. \cite{VidNun17} uses also the algorithm of \cite{NagEdw14}, added with variable gains, for exact tracking in MIMO systems, with known control matrix. In \cite{VidNun16}, an extension of \cite{NagEdw14} is worked out, assuming that the control matrix is \emph{uncertain} but \emph{constant, symmetric and positive definite}. The fragility of the case with non-symmetric uncertain input matrix is further studied in \cite{KeiNun19}, finding that lack of symmetry leads easily to instability.

Recently, in \cite{MorRio21pre} a further extension has been obtained. Quasi-continuous plus linear feedback terms are used to establish finite-time convergence in the presence of \emph{time-varying and state dependent uncertain control matrix} and uncertainties and perturbations. Again a non-smooth Lyapunov function is used, and the stability result requires the control matrix to have positive definite symmetric part. Note that most of the previous MIMO versions of the Super-Twisting have \emph{scalar gains}, strongly limiting the possibilities for control design. 

In this paper our aim is to further contribute to a better understanding of the MIMO design of the Super-Twisting algorithm. %
We use a smooth homogeneous Lyapunov function, motivated by a passivity interpretation, to design a MIMO Super-Twisting algorithm, with discontinuous feedback terms, in contrast to the usual quasi-continuous ones of the previous MIMO algorithms. The control matrix is assumed to be time and state dependent, with an \emph{uncertain but constant left} matrix factor. Sufficient convergence conditions are established for \emph{full matrix gains} instead of the accustomed scalar ones. Since the coupling in the control terms can be designed, contrary to the quasi-continuous class of control laws, it is in principle possible to take advantage of this design freedom to partially compensate for the natural coupling of the plant's channels caused by the uncertainty. Moreover, we consider not only discontinuous versions of the Super-Twisting algorithm, but also a full family of continuous and homogeneous approximations, including a linear PI-control law.

\subsection{Notation and preliminaries}

\label{subsec:Notation-and-preliminaries}We recall here some classes
of matrices (see e.g. \cite{HorJoh91}). A matrix $A=\left[a_{ij}\right]\in\mathbb{R}^{n\times n}$
is said to be \emph{positive definite} if it is symmetric and the
quadratic form $x^{T}Ax>0$ for all $x\neq0$. It is \emph{positive
quasi-definite} if $A+A^{T}$ is positive definite. $A$ is said to
be \emph{positive stable} if all its eigenvalues have positive real
parts. $A$ is a \emph{Z-matrix} if $a_{ij}\leq0$
for $i\neq j$. $A$ is called an \emph{M-Matrix} if $A$ is a Z-matrix
and is positive stable. $A$ is said to be \emph{strictly row diagonally dominant} if $\left|a_{ii}\right|>\sum_{j\neq i}\left|a_{ij}\right|$, for $i=1,\cdots,n$. %
It is \emph{strictly column diagonally dominant} if $A^{T}$ is strictly
row diagonally dominant.

Recall also the concept of weighted-homogeneous functions (see e.g. \cite{BacRos05}). %
Fix a set of coordinates $x=[x_1,\ldots,x_n]^\top \in \mathbb{R}^n$, and the \emph{coordinate weights} $\textbf{r}\triangleq[r_1,...,r_n]^{\top}\in \mathbb{R}_{>0}^n$.
Define the family of dilations $\Delta_{\epsilon}^{\bf r}$ such that $\Delta_{\epsilon}^{\mathbf{r}} x \triangleq[\epsilon^{r_{1}}x_{1},...,\epsilon^{r_n}x_n]^\top$.
A function $V:\mathbb{R}^n\to \mathbb{R}$ (resp. a vector field $f:\mathbb{R}^{n} \rightarrow \mathbb{R}^{n}$) is said to be $\mathbf{r}$-homogeneous of degree $l\in \mathbb{R}$, 
if for all $ \epsilon>0$ and all $ x \in \mathbb{R}^n$ the equality $V(\Delta_{\epsilon}^{\bf r}{x})=\epsilon^{l}V({x})$  (resp. $f(\Delta_{\epsilon}^{\bf r}x)=\epsilon^{l}\Delta_{\epsilon}^{\bf r}f(x)$) holds. 

We use the following notation: If $x\in\mathbb{R}$ and
$d\in\mathbb{R}$, $\left\lceil x\right\rfloor ^{d}=\left|x\right|^{d}\mbox{{\rm sign}}\left(x\right)$
is the signed power $d$ of $x$. If $x\in\mathbb{R}^{n}$ is a vector, 
functions $\left\lceil x\right\rfloor ^{d}$ and $\left|x\right|^{d}$
will be applied element-wise, e.g. $\left\lceil x^{T}\right\rfloor ^{d}=\left[\left\lceil x_{1}\right\rfloor ^{d},\cdots,\left\lceil x_{n}\right\rfloor ^{d}\right]$. %
The same notation is also used for matrices. $\mathbf{1}_{n}\in\mathbb{R}^{n}$ is a vector of ones.

\section{Problem formulation and proposed solution}

\label{sec:Problem-formulation}
Consider a MIMO uncertain system,
given by
\begin{align}
\dot{x} & =G\left(t,x\right)u+\delta\left(t\right)\,,\label{eq:Plant}
\end{align}
where $x\in\mathbb{R}^{m}$ is the state, $u\in\mathbb{R}^{m}$ is
the control input, $\delta\left(t\right)\in\mathbb{R}^{m}$ is a matched
perturbation and $G\left(t,x\right)\in\mathbb{R}^{m\times m}$ is
the continuous (uncertain) control matrix, i.e. $G \in \mathcal{G}$, where $\mathcal{G}$ is a family of matrix functions. \eqref{eq:Plant} results e.g. when a sliding mode controller has to be designed \cite{ShtEdw14,Kha02}.

In particular, we assume that $G \in \mathcal{G}$ can be represented as
\begin{equation}
G\left(t,x\right)=\Upsilon G_{0}\left(t,x\right),\label{eq:UncertainG}
\end{equation}
where $G_{0}\left(t,x\right)\in\mathbb{R}^{m\times m}$ is a \emph{known} regular
 matrix, i.e. $\left| \det G_{0}\left(t,x\right) \right| \geq \varepsilon > 0$, $\forall \left(t,x\right)$, %
 and $\Upsilon=\left[\upsilon_{ij}\right]\in\mathbb{R}^{m\times m}$
is constant and regular but \emph{uncertain}. This implies that $\det G\left(t,x\right)\neq0$
for all $\left(t,x\right)$ and that system \eqref{eq:Plant} has vector relative degree $\mathbf{1}_{m}$. 
We also assume that $0<\gamma_{1}\leq\left\Vert G_{0}\left(t,x\right)\right\Vert \leq\gamma_{2}$
for all $\left(t,x\right)$, for some positive constants. Assume
further that the values of $\Upsilon$ are described by the following
inequalities
\begin{equation}
\begin{split}
\upsilon_{ii} & \geq\underline{\upsilon}_{ii}>0,\quad\forall\quad i=1,...,m,\\
\left|\upsilon_{ij}\right| & \leq\bar{\upsilon}_{ij}\quad\forall\quad i=1,...,m,j=1,...,m\,.
\end{split}
\label{eq:UncertaintyUpsilon}
\end{equation}
Note that each $\Upsilon$ satisfying \eqref{eq:UncertaintyUpsilon}
has positive diagonal entries. The perturbation vector signal $\delta\left(t\right)$
is supposed to be a Lipschitz function of time, i.e. there is a constant
$C\geq0$ such that its derivative (where it exists) is bounded 
\begin{equation}
\left\Vert \dot{\delta}\left(t\right)\right\Vert \leq C,\,\forall t\geq0\,.\label{eq:BoundDelta}
\end{equation}

The \emph{objective} is to design a control law for $u$, such that
$u\left(t\right)$ is a continuous function of time and renders $x=0$
robustly asymptotically stable, despite the perturbations $\delta\left(t\right)$
and the uncertain control matrix.

\subsection{Proposed control structure}

We propose the control law with integral term given by
\begin{equation}
\begin{split}
u & =-K\left(t,x\right)\left\lceil x\right\rfloor ^{\frac{1}{1-l}}+B\left(t,x\right)v \\
\dot{v} & =-K_{I}\left(t,x\right)\left\lceil x\right\rfloor ^{\frac{1+l}{1-l}}\,,
\end{split}
\label{eq:PI-Control}
\end{equation}
where $K\left(t,x\right)\in\mathbb{R}^{m\times m}$, $K_{I}\left(t,x\right)\in\mathbb{R}^{m\times m}$,
and $B\left(t,x\right)\in\mathbb{R}^{m\times m}$ are (continuous)
and regular gain matrices to be designed, and $l$, the degree of
homogeneity, is a freely chosen parameter, taking values in the interval
$l\in\left[-1,0\right]$. We also require 
$a_{1}\leq\left\Vert K\left(t,x\right)\right\Vert \leq a_{2}$, 
$a_{3}\leq\left\Vert K_{I}\left(t,x\right)\right\Vert \leq a_{4}$, 
$a_{5}\leq\left\Vert B\left(t,x\right)\right\Vert \leq a_{6}$
for all $\left(t,x\right)$, for some constants $a_i>0$.

This control law is made entirely of continuous terms for $l \in (-1,0]$ and in $l = -1$ the only discontinuous term is behind the integrator, 
making $u(t)$ a continuous signal $\forall l \in [-1, 0]$.

The closed-loop system \eqref{eq:Plant}, \eqref{eq:PI-Control}, 
is given by
\begin{equation}
\begin{split}
\dot{x} & =G\left(t,x\right)\left(-K\left(t,x\right)\left\lceil x\right\rfloor ^{\frac{1}{1-l}}+B\left(t,x\right)x_{I}\right)\,,\\
\dot{x}_{I} & =-K_{I}\left(t,x\right)\left\lceil x\right\rfloor ^{\frac{1+l}{1-l}}+\Delta\left(t,x,\dot{x}\right)\,,
\end{split}
\label{eq:Closed-Loop}
\end{equation}
where $x_{I} =v+B^{-1}\left(t,x\right)G^{-1}\left(t,x\right)\delta\left(t\right)$ and
\begin{align}
\Delta\left(t,x,\dot{x}\right) & =\frac{d}{dt}\left(B^{-1}\left(t,x\right)G^{-1}\left(t,x\right)\delta\left(t\right)\right)\,.\label{eq:DefDelta}
\end{align}

\begin{rem}
\label{rem:On Delta}The definition of $\Delta$ in \eqref{eq:DefDelta}
raises immediately the issue, that in general $\Delta$ depends on
$\left(t,x\right)$ and $\dot{x}$, even when $\delta\left(t\right)$
is assumed only to be a function of time. Since $\dot{x}$ depends
on $u$ we get an algebraic loop. In order to avoid this, we will
require $G\left(t,x\right)B\left(t,x\right)$ to be \emph{constant}, that is,
\begin{equation}
G\left(t,x\right)B\left(t,x\right)=\mathcal{N},\label{eq:DesignCond5}
\end{equation}
where $\mathcal{N}\in\mathbb{R}^{m\times m}$ is a constant and regular
matrix. If \eqref{eq:DesignCond5} is satisfied, then $\Delta  =\frac{d}{dt}\left(\mathcal{N}^{-1}\delta\left(t\right)\right)=\mathcal{N}^{-1}\dot{\delta}\left(t\right)$.
In this case, the boundedness of $\dot{\delta}\left(t\right)$ \eqref{eq:BoundDelta}
implies the boundedness of $\Delta$.
\end{rem}
\begin{rem}
For $l=-1$ system \eqref{eq:Closed-Loop}
has a discontinuous right-hand side, and the solutions are
understood in the sense of Filippov \cite{Fil88}. In the scalar case, $m=1, l=-1$, the classic Super-Twisting Algorithm \cite{Lev93} is obtained, and thus \eqref{eq:PI-Control} is a MIMO generalization. \eqref{eq:Closed-Loop} has a switching surface at $x=0$, in contrast to the quasi-continuous MIMO versions in \cite{NagEdw14,VidNun16,LopMor19,MorRio21pre}, having a discontinuity only at $\left( x, x_{I} \right)=0$. For $l \in \left(-1, 0 \right]$ \eqref{eq:PI-Control} can be seen as a family of continuous \emph{approximations} of the discontinuous Super-Twisting Algorithm. For $l=0$ it becomes a linear PI-controller. 
\end{rem}

\subsection{Existence conditions for the controller}

The main result of the paper are the following sufficient conditions
for the existence of control gains solving the control problem. For
this we define the \emph{boundary matrix} $\mathcal{B}\left(\Upsilon\right)=\left[\upsilon_{b,ij}\right]\in\mathbb{R}^{m\times m}$,
corresponding to the interval \eqref{eq:UncertaintyUpsilon}, as $\upsilon_{b,ij} = -\bar{\upsilon}_{ij}$, if $i \neq j$, and $\upsilon_{b,ij} = \underline{\upsilon}_{ii}$, if $i=j$. %

\begin{thm}
\label{thm:Main}Consider system \eqref{eq:Plant}, 
satisfying assumptions \eqref{eq:UncertainG}, \eqref{eq:UncertaintyUpsilon}
and \eqref{eq:BoundDelta}. Under these conditions, if
the boundary matrix $\mathcal{B}\left(\Upsilon\right)$ is an M-matrix,
there exist gain matrices $K\left(t,x\right)$, $B\left(t,x\right)$
and $K_{I}\left(t,x\right)$ such that the control law \eqref{eq:PI-Control}
renders the origin $\left(x,x_{I}\right)=0$ of the closed-loop system
\eqref{eq:Closed-Loop}:
\begin{enumerate}
\item Finite-time stable for any $C>0$, if $l=-1$.
\item Finite-time stable for $C=0$, if $l\in\left(-1,0\right)$.
\item Exponentially stable for $C=0$, if $l=0$.
\end{enumerate}
Moreover, if $l\in\left(-1,0\right]$ and $C>0$, the closed-loop system \eqref{eq:Closed-Loop} is Input-to-State
Stable (ISS) with respect to the input $\Delta$. 
\end{thm}

The proof of Theorem \ref{thm:Main} is given in Section \ref{sec:Stability-analysis}
by using a strong Lyapunov function, motivated by a passivity interpretation.

Note that when $l=-1$ the proposed controller \eqref{eq:PI-Control}
has a discontinuous integral term and it is able to fully compensate
for bounded perturbations $\Delta$ \eqref{eq:DefDelta}, while for
$l\in\left(-1,0\right]$ the integral term is continuous but stability
will be attained only when $\Delta\equiv0$. This is well-known for
classic PI-controllers. In any case assuring stability of $\left(x,x_{I}\right)=0$,
the integral term is able to provide an estimation of the ``perturbation'', i.e. $v\left(t\right)\rightarrow-B^{-1}\left(t,x\right)G^{-1}\left(t,x\right)\delta\left(t\right)$
asymptotically (for $l=0$) or in finite time (for $l\in\left[-1,0\right)$). As required in the problem formulation, the control signal $u(t)$ is a continuous function, although for $l=-1$ the feedback is discontinuous.

\subsection{Gain scaling}

For a given set of appropriate gains and perturbation $\left(K,B,K_{I},\Delta\right)$
stabilizing the closed-loop system, it is possible to obtain a new
stabilizing set by introducing a gain scaling. First select \emph{time} and \emph{perturbation} scaling factors $T>0$ and $\kappa > 0$, respectively.  %
Then, we consider two possible gain and perturbation scalings \eqref{eq:GainScaling1} and \eqref{eq:GainScaling2}:
\begin{equation}
\left(K,B,K_{I},\Delta\right)\rightarrow\left(LK,B,L^{2}K_{I}, \kappa \Delta\right),\, L^{(1-l)}=\frac{T^{\left(1+l\right)}}{\kappa^{l}} \label{eq:GainScaling1}
\end{equation}
\begin{equation}
\left(K,B,K_{I},\Delta\right)\rightarrow\left(LK,LB,LK_{I}, \kappa \Delta\right),\, L=\frac{T^{\left(1+l\right)}}{\kappa^{l}}.\label{eq:GainScaling2}
\end{equation}
%
The change of coordinates in state and time, given by 
\[
x\rightarrow \frac{T^{2}}{\kappa} x,\,x_{I}\rightarrow \frac{T}{\kappa} x_{I},\,t\rightarrow Tt,
\]
\[
x\rightarrow \left(\frac{T}{\kappa}\right)^{1-l} x,\,x_{I}\rightarrow \frac{T}{\kappa} x_{I},\,t\rightarrow Tt,
\]
for \eqref{eq:GainScaling1} and \eqref{eq:GainScaling2}, respectively, transforms the gain scaled system into \eqref{eq:Closed-Loop}, 
so that they are equivalent. 

And therefore, the gain scaled system
can compensate a perturbation $\kappa$ times larger, and it accelerates
the convergence by $T$. For \eqref{eq:GainScaling1} the selected $T$ and
$\kappa$ have to be constant, while for \eqref{eq:GainScaling2} they can
be functions of $\left(t,x\right)$. In particular, if $\kappa\left(t,x\right)\geq\bar{\kappa}>0$
is a continuous function, uniformly bounded by a constant $\bar{\kappa}>0$,
and if $T\left(t,x\right)=\bar{T}\kappa\left(t,x\right)>0$ with a
constant $\bar{T}>0$, then the transformation above is valid. The gain scaling becomes $L\left(t,x\right)=\kappa \left(t,x\right)\bar{T}^{(1+l)}$.
This allows, for example, to deal with a perturbation $\Delta$ with
a time and/or state dependent but known upper bound.

\subsection{Gain Design}
\label{subsec:Gain-Design}
Consider the following particular gain selection:
\begin{equation}
B\left(t,x\right)  =G_{0}^{-1}\left(t,x\right)\mathbb{B}\,,
K\left(t,x\right)  =G_{0}^{-1}\left(t,x\right)\mathcal{K}\,,\label{eq:DesignK}
\end{equation}
with $K_{I}\left(t,x\right)$, $\mathbb{B}$ and $\mathcal{K}$ constant $\mathbb{R}^{m\times m}$ matrices. %

For this selection, it is shown in 
Sections \ref{sub:weakLF} and \ref{sub:StrongLF} that Theorem \ref{thm:Main} is true if there exist constant and regular matrices $\Gamma\left(\Upsilon\right),M\left(\Upsilon\right),P\left(\Upsilon\right),\mathcal{N}\left(\Upsilon\right)\in\mathbb{R}^{m\times m}$, with $P$ diagonal and positive definite, and $\Gamma$ positive definite, such that the following relations are satisfied
\begin{align}
\Upsilon\mathbb{B} & =\mathcal{N}\,, \label{eq:DesignCond5_N}\\
\left(P\Upsilon\mathbb{B}\right)\Gamma^{-1}+\Gamma^{-1} \left(P\Upsilon\mathbb{B}\right)^{T} & =K_{I}+K_{I}^{T}\,,\label{eq:ALE}
 \\
 \left\lceil x^{T}\right\rfloor ^{1+l}P\Upsilon\mathcal{K}x & >0,\,\forall x\neq0\,,\label{eq:RemainDesignCond3}
 \\
z^{T}M\Upsilon\mathbb{B}z & >0,\,\forall z\neq0\,. \label{eq:DesignCond4_N}
\end{align}
Choosing $\mathcal{N}=\Upsilon\mathbb{B}$ and $M=\left(\Upsilon\mathbb{B}\right)^{T}$ immediately solve \eqref{eq:DesignCond5_N} and \eqref{eq:DesignCond4_N}, respectively.  \eqref{eq:ALE} is an Algebraic Lyapunov Equation for $\Gamma^{-1}$. For every \textit{positive diagonal} $\mathbb{B}$, is $P\Upsilon\mathbb{B}$ positive stable (see Section \ref{sub:Feasibility}), and therefore for every \textit{positive quasi-definite} $K_{I}$ there is a unique positive definite solution $\Gamma^{-1}$ of \eqref{eq:ALE}. Finally, $P$ and $\mathcal{K}$ are selected such that \eqref{eq:RemainDesignCond3} is fulfilled. Lemma \ref{lem:HomForms} (Section \ref{sub:Feasibility}) shows that this is always possible, and that $\mathcal{K}$ depends on the bounds \eqref{eq:UncertaintyUpsilon} of $\Upsilon$. In particular, if $l=-1$, \eqref{eq:RemainDesignCond3} can be met for any \textit{positive diagonal} $\mathcal{K}$. 

For $l=-1$, in conclusion, stability is assured (for some $C$) for arbitrary $\mathbb{B}$ and $\mathcal{K}$ positive diagonal and an arbitrary positive quasi-definite $K_{I}$.

To tune the velocity and attain any size $C>0$ \eqref{eq:BoundDelta} of the perturbation, the gain scalings \eqref{eq:GainScaling1} or \eqref{eq:GainScaling2} are used. %

One interesting feature of our result, in contrast to most previous
ones in e.g. \cite{NagEdw14,VidNun16,MorRio21pre} that use scalar gains, is that conditions \eqref{eq:DesignCond5_N}-\eqref{eq:DesignCond4_N} characterize full gain matrices. %
Of course, it would be of great interest to describe the full set of solutions
of these design relations.

\subsection{Discussion of the results}
Note that for the previous particular selection of the gain matrices, the
closed-loop dynamics become 
\[
\dot{x} =\Upsilon\left(-\mathcal{K}\left\lceil x\right\rfloor ^{\frac{1}{1-l}}+\mathbb{B}x_{I}\right),\,
\dot{x}_{I}  =-K_{I}\left\lceil x\right\rfloor ^{\frac{1+l}{1-l}}+\Delta\left(t\right),
\]
which, in the absence of $\Delta$, is a homogeneous system of degree
$d_{c}=l$, with weight vectors $\mathbf{r}_{1}=\left(1-l\right)\mathbf{1}_{m}$
and $\mathbf{r}_{2}=\mathbf{1}_{m}$ for the vectors $x$ and $x_{I}$, respectively.


M-Matrices are frequently used in the study of interconnected systems (see e.g. \cite[Ch. 9]{Kha02}). Since the input matrix, and thus $\Upsilon$, represents interconnections between the different states, it seems natural that $\mathcal{B}\left(\Upsilon\right)$ should be an M-matrix.

Note also that, to be concrete, we have chosen to characterize the set of constant uncertain matrices $\Upsilon$ as an interval matrix \eqref{eq:UncertaintyUpsilon}. But other specifications can be used.

System  \eqref{eq:Closed-Loop} can be seen as the feedback interconnection of a passive system with a (passive) integrator, and thus it is also a passive system. The storage function \eqref{eq:WeakLF} (Section \ref{sub:weakLF}) is the sum of the two individual storage functions, and is, as usual, a weak Lyapunov function for \eqref{eq:Closed-Loop}. Although this interpretation leads to the strong Lyapunov function used for the proof in Section \ref{sec:Stability-analysis}, it also imposes a strong restriction on the control matrix $G$, given by \eqref{eq:DesignCond2}. %
It leads in turn to the requirement, that the uncertainty in the control matrix $G$ has to be parametrized by a \emph{constant} matrix $\Upsilon$ \eqref{eq:UncertainG}, instead of more general uncertainties as in \cite{MorRio21pre}. \eqref{eq:UncertainG} is the tightest condition of the paper. Although \eqref{eq:DesignCond5} is also very restrictive, if \eqref{eq:UncertainG} is fulfilled \eqref{eq:DesignCond5} is easily satisfied. Relaxing \eqref{eq:UncertainG} requires a different class of Lyapunov functions.

\section{Numerical Example}

Consider the same control problem as in \cite{KeiNun19} of the uncertain dynamics of a planar manipulator monitored by a fixed camera with optical axis orthogonal to the robot workspace plane with negligible manipulator dynamics,
\begin{equation}
\dot{\sigma} = G\left(\theta \right) u + \delta(t), \; G\left(\theta \right) = 
		\begin{bmatrix}
			\cos (\theta) & \sin (\theta) \\
			-\sin (\theta) & \cos (\theta)
		\end{bmatrix},
\end{equation}
where $\sigma = [x \: y]^T \in \mathbb{R}^2$ is the position in the $(x, y)$ plane, $\theta$ is the (constant) rotation angle of the camera. For an uncalibrated camera $G\left(\theta \right) = \Upsilon G\left(\bar{\theta} \right)$, for some nominal value $\bar{\theta}$ of $\theta$ and $\Upsilon= G\left(\Delta\theta \right)$, $\Delta\theta = \theta - \bar{\theta}$.

The analysis in \cite{KeiNun19} shows that the proposed unitary super-twisting can handle $|\Delta| < 19.6^{\circ}$ before instability. Since $\mathcal{B}\left( G\left(\Delta\theta \right) \right)$ is an M-matrix for $|\Delta| < 45^{\circ}$, our analysis assures stability for this interval.

\begin{figure}
\centerline{\includegraphics[scale=1.2]{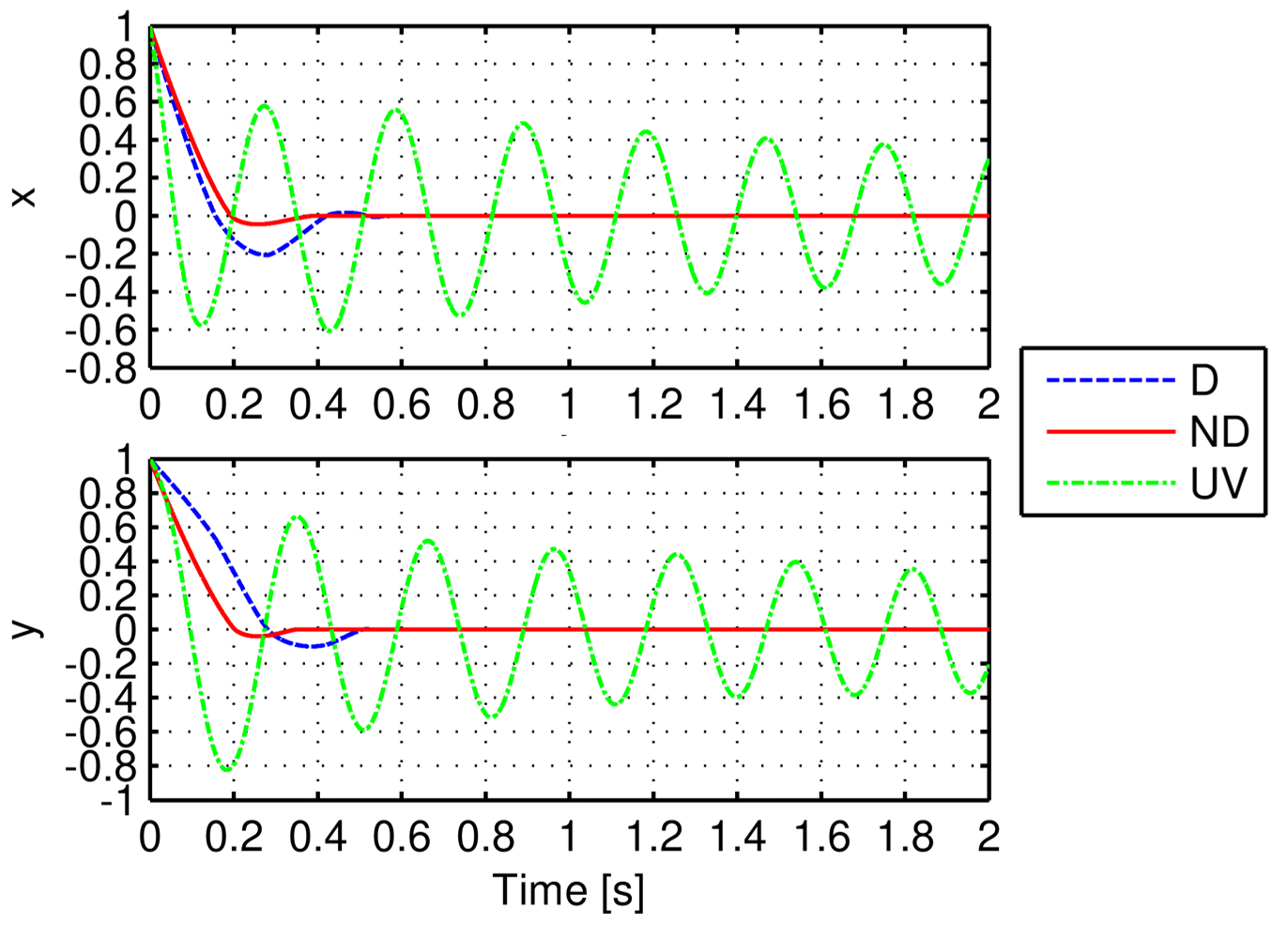}}
\caption{Origin stabilization for the controller with non-diagonal gains (ND), the one with diagonal gains (D) and the unit-vector-like (UV).}
\label{fig1}
\end{figure}

For the shake of comparison, Fig. \ref{fig1} shows some simulations of the closed-loop with three controllers: (i) controller \eqref{eq:PI-Control}, selecting the gains as not diagonal matrices (ND)
\begin{equation}
\mathcal{K} = \begin{bmatrix}
5 & -2.4 \\ 2.6 & 6
\end{bmatrix}, \: \: K_I = \begin{bmatrix}
10 & -6.4 \\ 8 & 12
\end{bmatrix}, \: \: \mathbb{B} = \begin{bmatrix}
1 & 0 \\ 0 & 1
\end{bmatrix},
\end{equation}
(ii) controller \eqref{eq:PI-Control} with the diagonal part of the previous gain matrices, and (iii) the unit-vector-like (UV) controller 
\[
u = -k_{1} \frac{\sigma}{\| \sigma \| ^\frac{1}{2}} - k_{2} \sigma + \xi , \; \dot{\xi} = -k_{3} \frac{\sigma}{\| \sigma \|} - k_{4} \sigma 
\] 
with the scalar gains proposed in \cite{KeiNun19}. %
All simulations were performed with $l=-1$, $\Delta \theta = 30 ^{\circ}$, $\delta = [\sin 5 t, \sin  2 t]^T$ and the ode1 Euler solver in Simulink with a fixed step of $1 \times 10 ^{-4}$.


\section{Stability analysis}

\label{sec:Stability-analysis}To prove Theorem \ref{thm:Main}, a Lyapunov function is developed in two steps: 
(i) A weak Lyapunov function is proposed. %
(ii) Adding a cross-term, a strong Lyapunov function is derived. 

\subsection{Special matrices and homogeneous functions }

\label{subsec:Some-homogeneous-forms}We collect here some useful
results on matrices.

\begin{lem}
\cite[Theorem 2.3 (M35,H24,G20)]{BerPle94}\label{lem:GeneralProperties}
Let $A=\left[a_{ij}\right]\in\mathbb{R}^{n\times n}$. Then, the following
implications hold:
$(C1)\Rightarrow(C2)\Rightarrow(C3)\,$.
If $A$ is a Z-matrix all conditions are equivalent to $A$ is a nonsingular
M-matrix.
\begin{enumerate}
\item[(C1)] $A$ has all positive diagonal elements and there exists a positive
diagonal matrix $D$ such that $AD$ is strictly row diagonally dominant, that is, %
$a_{ii}d_{i}>\sum_{j\neq i}\left|a_{ij}\right|d_{j},\,i=1,\cdots,n$.
\item[(C2)] There exists a positive diagonal matrix $D$ such that $AD+DA^{T}$
is positive definite.
\item[(C3)] $A$ is positive stable, that is, the real part of each eigenvalue
of $A$ is positive.
\end{enumerate}
\end{lem}
If $A$ is an M-matrix and $D$ is a positive diagonal matrix, then
$DA$ and $AD$ are M-matrices. Moreover, a symmetric M-matrix is
positive definite. The \emph{comparison matrix} of $A\in\mathbb{R}^{n\times n}$,
denoted by $\mathcal{M}\left(A\right)=\left[m_{ij}\right]\in\mathbb{R}^{n\times n}$,
is defined by $m_{ij}=-\left|a_{ij}\right|$ if $i\neq j$, and $m_{ij}=\left|a_{ij}\right|$ if $i=j$.
%
\begin{lem}
\cite[Theorem 2.5.3 item 2.5.3.15]{HorJoh91}\label{lem:MMatrixYSRDDM}
If $\Gamma$ is an M-matrix with diagonal positive entries, there
exist positive diagonal matrices $B$ and $\mathcal{D}$ such that
$B\Gamma\mathcal{D}$, and also $\mathcal{D}\Gamma^{T}B$, are both
strictly row diagonally dominant and strictly column diagonally dominant.
\end{lem}
Recall the following well-known property of continuous homogeneous
functions 
\begin{lem}
\label{lem:Homogeneity_Property}\cite{Hes66,MorCru20C}
Let $\eta:\mathbb{R}^{n}\rightarrow\mathbb{R}$ and $\gamma:\mathbb{R}^{n}\rightarrow\mathbb{R}_{+}$
be two continuous homogeneous functions, with weights $\textbf{r}=(r_{1},...,r_{n})$
and degrees $m$, with $\gamma(x)\geq0$, such that 
$\{x\in\mathbb{R}^{n}\setminus\{0\}:\gamma(x)=0\}\subseteq\{x\in\mathbb{R}^{n}\setminus\{0\}:\eta(x)<0\}$ holds.
Then, there exists a real number $\lambda^{*}$ such that, for all
$\lambda>\lambda^{*}$, for all $x\in\mathbb{R}^{n}\setminus\{0\}$
and some $c>0$, $\eta(x)-\lambda\gamma(x)<-c\left\Vert x\right\Vert _{\textbf{r},p}^{m}$. 
\end{lem}

\subsection{A weak Lyapunov function}
\label{sub:weakLF}
Consider the positive definite scalar function
\begin{equation}
W\left(x,x_{I}\right)=\frac{1-l}{2}\left\lceil x^{T}\right\rfloor ^{\frac{1}{1-l}}P\left\lceil x\right\rfloor ^{\frac{1}{1-l}}+\frac{1}{2}x_{I}^{T}\Gamma x_{I}\,,\label{eq:WeakLF}
\end{equation}
where $P\in\mathbb{R}^{m\times m}$ is a \textit{constant, diagonal and positive
definite matrix}, 
and $\Gamma\in\mathbb{R}^{m\times m}$ is a \textit{constant and positive
definite matrix}. 
$W$ is continuously differentiable for $l\in\left(-1,0\right]$ and
Lipschitz continuous for $l=-1$. $W$ is homogeneous of degree $d_{W}=2$
with weight vectors $\mathbf{r}_{1}=\left(1-l\right)\mathbf{1}_{m}$
and $\mathbf{r}_{2}=\mathbf{1}_{m}$ for vectors $x$ and $x_{I}$,
respectively.

The derivative of $W$ along the trajectories of the closed-loop system
\eqref{eq:Closed-Loop} is given by (for simplicity, we omit the arguments)
\begin{align*}
\dot{W} & =-\left\lceil x^{T}\right\rfloor ^{\frac{1+l}{1-l}}PGK\left\lceil x\right\rfloor ^{\frac{1}{1-l}}\\
 & +\left\lceil x^{T}\right\rfloor ^{\frac{1+l}{1-l}}\left(PGB-K_{I}^{T}\Gamma\right)x_{I}+x_{I}^{T}\Gamma \Delta\,.
\end{align*}
If $\exists K,K_{I},B$ such that for each $G \in \mathcal{G}$ exist $P,\Gamma$ so that
\begin{align}
PG\left(t,x\right)B\left(t,x\right) & =K_{I}^{T}\left(t,x\right)\Gamma\,,\label{eq:DesignCond2}\\
\left\lceil x^{T}\right\rfloor ^{1+l}PG\left(t,x\right)K\left(t,x\right)x & >0,\,\forall x\neq0\,,\label{eq:DesignCond3}
\end{align}
are satisfied, and if $\Delta\left(t\right)\equiv0$,
it follows that for all $x\neq0$
\begin{align}
\dot{W} & =-\left\lceil x^{T}\right\rfloor ^{\frac{1+l}{1-l}}PG\left(t,x\right)K\left(t,x\right)\left\lceil x\right\rfloor ^{\frac{1}{1-l}}<0\,.\label{eq:DotWNeg}
\end{align}
This means that $W$ is a weak Lyapunov function. If the closed-loop
system is time-invariant, then Lasalle's invariance principle will
imply asymptotic stability, but not in the general case. %

\subsection{A strong Lyapunov function}
\label{sub:StrongLF}
We add a cross-term to $W$
\begin{equation}
V\left(x,x_{I}\right)=\frac{2}{2-l}\mu W^{\frac{2-l}{2}}\left(x,x_{I}\right)-x_{I}^{T}Mx\,,\label{eq:StrongLF}
\end{equation}
where $M\in\mathbb{R}^{m\times m}$ is a \textit{constant regular matrix} and
$\mu\in\mathbb{R}_{+}$ is a constant positive scalar. Since $\frac{2-l}{2}\geq1$,
$V$ is continuously differentiable for $l\in\left(-1,0\right]$ and
Lipschitz continuous for $l=-1$. $V$ is also homogeneous of degree
$d_{V}=2-l$ with weight vectors $\mathbf{r}_{1}=\left(1-l\right)\mathbf{1}_{m}$, $\mathbf{r}_{2}=\mathbf{1}_{m}$ for $x$ and $x_{I}$,
respectively. To render $V$ positive definite it is necessary to
select $\mu>0$ sufficiently large, what can be shown using Lemma
\ref{lem:Homogeneity_Property}.

The derivative of $V$ is given by
\begin{multline*}
\dot{V}  =-\mathcal{W}\left(t,x,x_{I}\right)+\rho\left(x,x_{I}\right)\Delta\,, \\
\mathcal{W}\left(t,x,x_{I}\right)  \triangleq\mu W^{\frac{-l}{2}}\left(x,x_{I}\right)\left\lceil x^{T}\right\rfloor ^{\frac{1+l}{1-l}}PGK\left\lceil x\right\rfloor ^{\frac{1}{1-l}}\\
  -x^{T}M^{T}K_{I}\left\lceil x\right\rfloor ^{\frac{1+l}{1-l}}-x_{I}^{T}MGK\left\lceil x\right\rfloor ^{\frac{1}{1-l}}
  +x_{I}^{T}MGBx_{I}\,,\\
\rho\left(x,x_{I}\right)  \triangleq W^{\frac{-l}{2}}\left(x,x_{I}\right)x_{I}^{T}\tilde{\Gamma}-x^{T}M^{T}\,,
\end{multline*}
where $\tilde{\Gamma} \triangleq\mu\Gamma$. 
Note that $\tilde{\Gamma}$ can be selected independently of $\mu$.
Although function $\mathcal{W}$ itself is not assured to be homogeneous
(due to the dependence on time and $x$ of $G$, and the matrix gains),
it can be bounded by a homogeneous function of degree $d_{\mathcal{W}}=2$,
while each component of the row vector $\rho\left(x,x_{I}\right)$ is homogeneous of
degree $d_{\rho}=1-l$.

Due to \eqref{eq:DesignCond3} the first term of $\mathcal{W}$ is
positive for $x\neq0$ and the latter one is positive for $x_{I}\neq0$
 if for each $G\left(t,x\right) \in \mathcal{G}$ there is $M$ so that condition 
\begin{align}
z^{T}MG\left(t,x\right)B\left(t,x\right)z & >0,\,\forall x,\forall z\neq0 \label{eq:DesignCond4}
\end{align}
is satisfied, i.e. $MG\left(t,x\right)B\left(t,x\right)$
is positive quasi-definite.

Since the first term is positive and vanishes on the set $\mathcal{S}=\left\{ \left(x,x_{I}\right)\in\mathbb{R}^{2m}|x=0\right\} $,
then
\begin{align*}
\left.\mathcal{W}\right|_{\mathcal{S}} & =x_{I}^{T}MG\left(t,0\right)B\left(t,0\right)x_{I}\,,
\end{align*}
which is positive for $x_{I}\neq0$ if condition \eqref{eq:DesignCond4}
is fulfilled. Using Lemma \ref{lem:Homogeneity_Property}\footnote{\label{fn:FootNote}Note that this Lemma has to be applied not directly
to function $\mathcal{W}$ but to its homogeneous bounding function.
We allow ourselves this lack of precision for simplicity of the presentation.} we conclude that $\mathcal{W}$ can be rendered positive definite
by selecting $\mu>0$ sufficiently large. In the absence of perturbation,
i.e. for $\delta\left(t\right)=0$, this implies the (uniform) global
asymptotic stability of the origin $\left(x,x_{I}\right)=0$ of \eqref{eq:Closed-Loop}.
Due to homogeneity of $V$ and of the positive definite bounding function
of $\mathcal{W}$, finite-time stability or exponential stability
is concluded. This proves items 2) and 3) of Theorem \ref{thm:Main}.

For the perturbed case, i.e. $\delta\left(t\right)\neq0$, condition
\eqref{eq:BoundDelta}, together with Remark \ref{rem:On Delta}, 
implies
that $\Delta$ is bounded, i.e. $\left\Vert \Delta\right\Vert \leq\tilde{C}$,
for some constant $\tilde{C}\geq0$. Thus, there is a constant
$\varrho\geq0$ such that
\[
\rho\left(x,x_{I}\right)\Delta\leq\varrho\tilde{C} V^{\frac{1-l}{2-l}}\left(x,x_{I} \right)\,.
\]
When $l\in\left(-1,0\right]$, $d_{\rho}<d_{\mathcal{W}}$, and therefore
near $\left(x,x_{I}\right)=0$ the term $\rho\left(t,x\right)$ dominates
and $\dot{V}$ cannot be rendered negative near zero. However, for
large values of $\left(x,x_{I}\right)$ it becomes negative. And using
standard Lyapunov arguments, we conclude that the system \eqref{eq:Closed-Loop}
is ISS with respect to the input $\Delta$. Or, equivalently, that
the trajectories of the closed-loop system are ultimately and uniformly
bounded \cite{Kha02}. This proves the last part of Theorem \ref{thm:Main}.

When $l=-1$, then $d_{\rho}=d_{\mathcal{W}}$ and $\dot{V}$ is negative
definite for $\tilde{C}$ sufficiently small. This can be shown again
using Lemma \ref{lem:Homogeneity_Property} (see again footnote \ref{fn:FootNote}).
This proves item 1) of Theorem \ref{thm:Main}. Note that the previous
argument assures stability only for a perturbation bound $\tilde{C}$
sufficiently small. Using the scaling of the gains, either \eqref{eq:GainScaling1}
or \eqref{eq:GainScaling2}, an arbitrary size of the perturbation
can be accommodated. Even a perturbation which grows with time and
the state, when \eqref{eq:GainScaling2} is used.

\subsection{Feasibility of conditions \eqref{eq:DesignCond5}, \eqref{eq:DesignCond2}, \eqref{eq:DesignCond3}, \eqref{eq:DesignCond4}}
\label{sub:Feasibility}
In Sections \ref{sub:weakLF} and \ref{sub:StrongLF} is shown that for $V$ \eqref{eq:StrongLF}
to be a strong Lyapunov function for \eqref{eq:Closed-Loop} conditions  
\eqref{eq:DesignCond2}, \eqref{eq:DesignCond3} and \eqref{eq:DesignCond4} have to be satisfied. However, these conditions are very general and it is difficult to determine if they are feasible, under general uncertainty conditions for $G$. We show here that this is the case under the hypothesis of Theorem \ref{thm:Main}. 

For this note first that, for the particular selection of gain matrices $K$, $B$, $K_{I}$ performed in Section \ref{subsec:Gain-Design}, these conditions become \eqref{eq:ALE}, \eqref{eq:RemainDesignCond3} and \eqref{eq:DesignCond4_N}, respectively. The extra condition \eqref{eq:DesignCond5} becomes \eqref{eq:DesignCond5_N}. An important property of $\Upsilon$ is stated first. 

\begin{lem}
\label{lem:UpsilonGSDD}Let $\Upsilon$ satisfy \eqref{eq:UncertaintyUpsilon}.
$\Upsilon$ satisfies (C1) in Lemma \ref{lem:GeneralProperties} if and only if the boundary matrix $\mathcal{B}\left(\Upsilon\right)$ %
is an M-matrix. 
\end{lem}
\begin{proof}
Since $\mathcal{B}\left(\Upsilon\right)$ satisfies \eqref{eq:UncertaintyUpsilon}
the necessity is clear. If $\mathcal{B}\left(\Upsilon\right)$ is
an M-matrix, then using \eqref{eq:UncertaintyUpsilon} and Lemma \ref{lem:GeneralProperties} there exist $d_{i}>0$
such that for each $i=1,\cdots,m$
the inequalities
\[
\upsilon_{ii}d_{i}\geq\underline{\upsilon}_{ii}d_{i}>\sum_{j\neq i}^{m}\overline{\upsilon}_{ij}d_{j}\geq\sum_{j\neq i}^{m}\upsilon_{ij}d_{j}\,,\,i=1,\cdots,m,
\]
are obtained, so that $\Upsilon$ satisfies (C1) in Lemma \ref{lem:GeneralProperties}. \qed
\end{proof}
The previous lemma is of great importance since it shows that all matrices $\Upsilon$  satisfying \eqref{eq:UncertaintyUpsilon} 
have generalized diagonal dominance with \emph{the same} $d_i$ that works for the boundary matrix. %

As already noted in Section \ref{subsec:Gain-Design} 
conditions \eqref{eq:DesignCond5_N} and \eqref{eq:DesignCond4_N} are easily fulfilled selecting $\mathcal{N}$ and $M$ appropriately. Consider \eqref{eq:ALE}, with constant $K_{I}$.  
By hypothesis of the Theorem, $\mathcal{B}\left(\Upsilon\right)$
is an M-matrix with positive diagonal entries. Lemma \ref{lem:UpsilonGSDD}
implies that each $\Upsilon$ satisfies (C1) in Lemma \ref{lem:GeneralProperties}, and therefore, it is positive
stable. Moreover, by \cite[Theorem 2.5.8]{HorJoh91} $P\Upsilon\mathbb{B}$
is positive stable for arbitrary positive diagonal matrices $P$ and
$\mathbb{B}$. Since \eqref{eq:ALE} is an Algebraic Lyapunov Equation
for $\Gamma^{-1}$, there is a unique, (symmetric) positive definite
$\Gamma^{-1}$ for any positive quasi-definite $K_{I}$, as required.


Finally, the next Lemma shows that if $\mathcal{B}\left(\Upsilon\right)$ %
is an M-matrix, then \eqref{eq:RemainDesignCond3} is fulfilled selecting appropriately $P$ and $\mathcal{K}$.
\begin{lem}
\label{lem:HomForms}Suppose that $\mathcal{B}\left(\Upsilon\right)$
is an M-matrix and that $l\in\left[-1,0\right]$. Then there is a
positive diagonal matrix $P=\mbox{{\rm diag}}\left\{ p_{i}\right\} $, 
and a matrix $\mathcal{\mathcal{K}}=\left[k_{ij}\right]\in\mathbb{R}^{m\times m}$,
with positive diagonal elements, i.e. $k_{ii}>0$, such that \eqref{eq:RemainDesignCond3}
is satisfied.
\end{lem}
\begin{proof}
We decompose matrix $\mathcal{K}$ as $\mathcal{K}=\mathcal{K}_{d}+\mathcal{K}_{o}$,
where $\mathcal{K}_{d}=\mbox{{\rm diag}}\left\{ k_{11},\cdots,k_{mm}\right\} $,
$k_{ii}>0$, is its diagonal part, and $\mathcal{K}_{o}=\mathcal{K}-\mathcal{K}_{d}$
contains the off-diagonal terms. For arbitrary $P$ and $\mathcal{K}_{d}$,
and a matrix $\Upsilon$ satisfying \eqref{eq:UncertaintyUpsilon},
and for any $l\geq-1$, the following inequalities are satisfied for all
$x\in\mathbb{R}^{m}$
\begin{multline*}
F_{d}\left(x \right) \triangleq \left\lceil x^{T} \right\rfloor ^{1+l} P\Upsilon\mathcal{K}_{d} x = \sum_{i=1}^{m}p_{i}\upsilon_{ii}k_{ii}\left|x_{i}\right|^{2+l} \\ +\sum_{i=1}^{m}\sum_{j\neq i}^{m}p_{i}\upsilon_{ij}k_{jj}\left\lceil x_{i}\right\rfloor ^{1+l} x_{j}\\
\geq\sum_{i=1}^{m}p_{i}\upsilon_{ii}k_{ii}\left|x_{i}\right|^{2+l}-\sum_{i=1}^{m}\sum_{j\neq i}^{m}p_{i}\left|\upsilon_{ij}\right|\left|k_{jj}\right| \left|x_{i}\right|^{1+l} \left|x_{j}\right|
\end{multline*}
Using Young's inequality for $l \geq -1$, i.e.
\[
\left|x_{i}\right|^{1+l}\left|x_{j}\right| \leq \frac{1+l}{2+l}\left|x_{i}\right|^{2+l}+\frac{1}{2+l}\left|x_{j}\right|^{2+l}\,,\,1+l \geq 0\,,
\]
leads to
\begin{multline*}
F_{d}\left(x \right) \geq \frac{1+l}{2+l} \sum_{i=1}^{m} \left|x_{i}\right|^{2+l} p_{i} \left( \upsilon_{ii}k_{ii} - \sum_{j\neq i}^{m} \left|\upsilon_{ij}\right|\left|k_{jj}\right|  \right) \\
+ \frac{1}{2+l} \sum_{i=1}^{m}p_{i} \left( \upsilon_{ii}k_{ii}\left|x_{i}\right|^{2+l} - \sum_{j\neq i}^{m} \left|\upsilon_{ij}\right|\left|k_{jj}\right|  \left|x_{j}\right|^{2+l} \right) \\
=\left(\left|x\right|^{2+l}\right)^{T}\left(\frac{1+l}{2+l}P\mathcal{M}\left(\Upsilon\right)\mathcal{K}_{d}+\frac{1}{2+l}\mathcal{K}_{d}\mathcal{M}^{T}\left(\Upsilon\right)P\right){\bf 1}\,.
\end{multline*}
Further, using \eqref{eq:UncertaintyUpsilon}, we obtain
\begin{align*}
F_{d}\left(x \right) & \geq \left|x^{T}\right|^{2+l} \mathcal{H}_{d}\left( \Upsilon \right) {\bf 1}\,, \\
\mathcal{H}_{d}\left( \Upsilon \right) &  \triangleq \frac{1+l}{2+l}P\mathcal{B}\left(\Upsilon\right)\mathcal{K}_{d}+\frac{1}{2+l}\mathcal{K}_{d}\mathcal{B}^{T}\left(\Upsilon\right)P \,.
\end{align*}
Similarly, using Young's inequality and \eqref{eq:UncertaintyUpsilon},
we obtain 
\begin{multline*}
F_{o}\left(x \right) \triangleq \left\lceil x^{T} \right\rfloor ^{1+l} P\Upsilon\mathcal{K}_{o}  x \leq \sum_{i=1}^{m}p_{i}\sum_{\theta\neq i}^{m}\left|\upsilon_{i\theta}\right|\left|k_{\theta i}\right|\left|x_{i}\right|^{2+l} \\
+\sum_{i=1}^{m}\sum_{j\neq i}^{m}p_{i}\sum_{\theta\neq j}^{m}\left|\upsilon_{i\theta}\right|\left|k_{\theta j}\right| \left|x_{i}\right|^{1+l} \left|x_{j}\right|
\leq \left|x^{T}\right|^{2+l} \mathcal{H}_{o}\left( \Upsilon \right) {\bf 1},
\end{multline*}
\begin{align*}
\mathcal{H}_{o}\left( \Upsilon \right) &  \triangleq \frac{1+l}{2+l}P\overline{\Upsilon}\left|\mathcal{K}_{o}\right|+\frac{1}{2+l}\left|\mathcal{K}_{o}\right|^{T}\overline{\Upsilon}^{T}P\,,
\end{align*}
and $\overline{\Upsilon}=\left[\overline{\upsilon}_{ij}\right]$.
Note that $\left|x^{T}\right|^{2+l} \mathcal{H}_{o}\left( \Upsilon \right) {\bf 1} \geq 0$. Finally,
\begin{align*}
x^{T}P\Upsilon\mathcal{K}\left\lceil x\right\rfloor ^{1+l} & \geq \left|x^{T}\right|^{2+l} \mathcal{H}_{d}\left( \Upsilon \right) {\bf 1} - \left|x^{T}\right|^{2+l} \mathcal{H}_{o}\left( \Upsilon \right) {\bf 1}\,.
\end{align*}
If $\mathcal{H}_{d}\left( \Upsilon \right)$ is strictly row diagonally dominant and it has positive diagonal entries
it follows from the previous inequalities that $F_{d}\left(x \right)>0$ for $x\neq0$. According to Lemma \ref{lem:MMatrixYSRDDM} if $\mathcal{B}\left(\Upsilon\right)$
is an M-matrix there exist positive diagonal matrices $P$ and $\mathcal{K}_{d}$
so that $P\mathcal{B}\left(\Upsilon\right)\mathcal{K}_{d}$, and therefore
$\mathcal{H}_{d}\left( \Upsilon \right)$,
are both strictly row and column diagonally dominant. If $l=-1$ this can be met choosing $P$ alone. Finally, selecting
the components of $\mathcal{K}_{o}$ sufficiently small, $x^{T}P\Upsilon\mathcal{K}\left\lceil x\right\rfloor ^{1+l}>0$
for all $x\neq0$. This happens if e.g. $\left|k_{ij}\right|\leq\epsilon$
for $i\neq j$, and $\epsilon \geq 0$ is sufficiently small. \qed
\end{proof}

\bibliographystyle{ieeetr}

\end{document}